\RequirePackage{ifpdf}
\ifpdf 
\documentclass[pdftex]{sigma}
\else
\documentclass{sigma}
\fi

\numberwithin{equation}{section}
\numberwithin{lemma}{section}
\numberwithin{remark}{section}

\begin{document}
\newcommand{\Si}{\Sigma}
\newcommand{\tr}{{\rm tr}}
\newcommand{\ad}{{\rm ad}}
\newcommand{\Ad}{{\rm Ad}}
\newcommand{\ti}[1]{\tilde{#1}}
\newcommand{\om}{\omega}
\newcommand{\Om}{\Omega}
\newcommand{\de}{\delta}
\newcommand{\al}{\alpha}
\newcommand{\te}{\theta}
\newcommand{\vth}{\vartheta}
\newcommand{\be}{\beta}
\newcommand{\la}{\lambda}
\newcommand{\La}{\Lambda}
\newcommand{\D}{\Delta}
\newcommand{\ve}{\varepsilon}
\newcommand{\ep}{\epsilon}
\newcommand{\vf}{\varphi}
\newcommand{\vfh}{\varphi^\hbar}
\newcommand{\vfe}{\varphi^\eta}
\newcommand{\fh}{\phi^\hbar}
\newcommand{\fe}{\phi^\eta}
\newcommand{\G}{\Gamma}
\newcommand{\ka}{\kappa}
\newcommand{\ip}{\hat{\upsilon}}
\newcommand{\Ip}{\hat{\Upsilon}}
\newcommand{\ga}{\gamma}
\newcommand{\ze}{\zeta}
\newcommand{\si}{\sigma}

\def\clA{\mathcal{A}}
\def\clC{\mathcal{C}}
\def\clD{\mathcal{D}}
\def\clF{\mathcal{F}}
\def\clG{\mathcal{G}}
\def\clH{\mathcal{H}}
\def\clK{\mathcal{K}}
\def\clI{\mathcal{I}}
\def\clM{\mathcal{M}}
\def\clR{\mathcal{R}}
\def\clU{\mathcal{U}}
\def\clO{\mathcal{O}}
\def\clQ{\mathcal{Q}}
\def\clL{\mathcal{L}}
\def\clN{\mathcal{N}}
\def\clS{\mathcal{S}}
\def\clT{\mathcal{T}}
\def\clY{\mathcal{Y}}
\def\clW{\mathcal{W}}
\def\clZ{\mathcal{Z}}

\def\bfa{{\bf a}}
\def\bfb{{\bf b}}
\def\bfc{{\bf c}}
\def\bfd{{\bf d}}
\def\bfe{{\bf e}}
\def\bff{{\bf f}}
\def\bfm{{\bf m}}
\def\bfn{{\bf n}}
\def\bfp{{\bf p}}
\def\bfu{{\bf u}}
\def\bfv{{\bf v}}
\def\bft{{\bf t}}
\def\bfx{{\bf x}}
\def\bfy{{\bf y}}
\def\bfg{{\bf g}}
\def\bfC{{\bf C}}
\def\bfA{{\bf A}}
\def\bfS{{\bf S}}
\def\bfJ{{\bf J}}
\def\bfI{{\bf I}}
\def\bfP{{\bf P}}
\def\bfr{{\bf r}}
\def\bfU{{\bf U}}
\def\bfE{{\bf E}}
\def\bfB{{\bf B}}
\def\bfPhi{{\bf \Phi}}
\def\bfphi{{\bf \phi}}

\def\bfal{\breve{\al}}
\def\bfbe{\breve{\be}}
\def\bfga{\breve{\ga}}
\def\bfnu{\breve{\nu}}
\def\bfsi{\breve{\sigma}}

\def\bPhi{\bar{\Phi}}

\def\hS{{\hat{S}}}
\def\ad{\rm ad}
\def\Ad{\rm Ad}

\newcommand{\li}{\lim_{n\rightarrow \infty}}
\def\mapright#1{\smash{
\mathop{\longrightarrow}\limits^{#1}}}

\newcommand{\mat}[4]{\left(\begin{array}{cc}{#1}&{#2}\\{#3}&{#4}
\end{array}\right)}
\newcommand{\thmat}[9]{\left(
\begin{array}{ccc}{#1}&{#2}&{#3}\\{#4}&{#5}&{#6}\\
{#7}&{#8}&{#9}
\end{array}\right)}

\newcommand{\p}{\partial}
\newcommand{\di}{{\rm diag}}
\newcommand{\oh}{\frac{1}{2}}
\newcommand{\su}{{\bf su_2}}
\newcommand{\uo}{{\bf u_1}}
\newcommand{\SL}{{\rm SL}(2,{\mathbb C})}
\newcommand{\GLN}{{\rm GL}(N,{\mathbb C})}
\newcommand{\PGLN}{{\rm PGL}(N,{\mathbb C})}

\def\SUN{{\rm SU}(N)}
\def\PSUN{{\rm PSU}(N)}
\def\sun{{\rm su}(N)}
\def\sln{{\rm sl}(N, {\mathbb C})}
\def\sl2{{\rm sl}(2, {\mathbb C})}
\def\SLN{{\rm SL}(N, {\mathbb C})}
\def\PSLN{{\rm PSL}(N, {\mathbb C})}
\def\SLT{{\rm SL}(2, {\mathbb C})}
\newcommand{\gln}{{\rm gl}(N, {\mathbb C})}
\newcommand{\PSL}{{\rm PSL}_2( {\mathbb Z})}
\def\f1#1{\frac{1}{#1}}
\def\lb{\lfloor}
\def\rb{\rfloor}
\def\sn{{\rm sn}}
\def\cn{{\rm cn}}
\def\dn{{\rm dn}}
\newcommand{\rar}{\rightarrow}
\newcommand{\upar}{\uparrow}
\newcommand{\sm}{\setminus}
\newcommand{\ms}{\mapsto}
\newcommand{\bp}{\bar{\partial}}
\newcommand{\bz}{\bar{z}}
\newcommand{\bw}{\bar{w}}
\newcommand{\bA}{\bar{A}}
\newcommand{\bL}{\bar{L}}
\newcommand{\btau}{\bar{\tau}}

\newcommand{\Sh}{\hat{S}}
\newcommand{\vtb}{\theta_{2}}
\newcommand{\vtc}{\theta_{3}}
\newcommand{\vtd}{\theta_{4}}

\def\mC{{\mathbb C}}
\def\mZ{{\mathbb Z}}
\def\mR{{\mathbb R}}
\def\mN{{\mathbb N}}
\def\mP{{\mathbb P}}
\def\mH{{\mathbb H}}

\def\frak{\mathfrak}
\def\gg{{\frak g}}
\def\gJ{{\frak J}}
\def\gS{{\frak S}}
\def\gL{{\frak L}}
\def\gG{{\frak G}}
\def\gk{{\frak k}}
\def\gK{{\frak K}}
\def\gl{{\frak l}}
\def\gh{{\frak h}}
\def\gH{{\frak H}}

\newcommand{\ran}{\rangle}
\newcommand{\lan}{\langle}
\def\f1#1{\frac{1}{#1}}
\def\lb{\lfloor}
\def\rb{\rfloor}
\newcommand{\slim}[2]{\sum\limits_{#1}^{#2}}

\allowdisplaybreaks

\renewcommand{\thefootnote}{$\star$}

\renewcommand{\PaperNumber}{065}

\FirstPageHeading

\ShortArticleName{Monopoles and Modif\/ications of Bundles over Elliptic Curves}

\ArticleName{Monopoles and Modif\/ications of Bundles\\ over Elliptic Curves\footnote{This paper is a contribution to the Proceedings of the Workshop ``Elliptic Integrable Systems, Isomonodromy Problems, and Hypergeometric Functions'' (July 21--25, 2008, MPIM, Bonn, Germany). The full collection
is available at
\href{http://www.emis.de/journals/SIGMA/Elliptic-Integrable-Systems.html}{http://www.emis.de/journals/SIGMA/Elliptic-Integrable-Systems.html}}}

\Author{Andrey M.~LEVIN~$^{\dag\ddag}$, Mikhail A.~OLSHANETSKY~$^{\dag\S}$ and Andrei V.~ZOTOV~$^{\dag\S}$}

\AuthorNameForHeading{A.M.~Levin, M.A.~Olshanetsky and A.V.~Zotov}

\Address{$^\dag$~Max Planck Institute of Mathematics, Bonn, Germany}

\Address{$^\ddag$~Institute of Oceanology, Moscow, Russia}
\EmailD{\href{mailto:alevin@wave.sio.rssi.ru}{alevin@wave.sio.rssi.ru}}

\Address{$^\S$~Institute of Theoretical and Experimental Physics, Moscow, Russia}
\EmailD{\href{mailto:olshanet@itep.ru}{olshanet@itep.ru}, \href{mailto:zotov@itep.ru}{zotov@itep.ru}}

\ArticleDates{Received November 20, 2008, in f\/inal form June 10,
2009; Published online June 25, 2009}

\Abstract{Modif\/ications of bundles over complex curves is an
operation that allows one to construct a new bundle from a given
one. Modif\/ications can change a topological type of bundle. We
describe the topological type in terms of the characteristic classes
of the bundle. Being applied to the Higgs bundles modif\/ications
establish an equivalence between dif\/ferent classical integrable
systems. Following Kapustin and Witten we def\/ine the modif\/ications
in terms of monopole solutions of the Bogomolny equation. We f\/ind
the Dirac monopole solution in the case $R$ $\times$ (elliptic curve). This
solution is a three-dimensional generalization of the Kronecker
series. We give two representations for this solution and derive a
functional equation for it generalizing the Kronecker results. We
use it to def\/ine Abelian modif\/ications for bundles of arbitrary
rank. We also describe non-Abelian modif\/ications in terms of
theta-functions with characteristic.}

\Keywords{integrable systems; f\/ield theory; characteristic classes}

\Classification{14H70; 14F05; 33E05; 37K20; 81R12}

\renewcommand{\thefootnote}{\arabic{footnote}}
\setcounter{footnote}{0}

\section{Introduction}

The modif\/ications (or the Hecke transformation) of bundles over
complex curves is a correspondence between two bundles $E$ and $\ti
E$. It is isomorphism in a complement of some divisor.  A~modif\/ication can change the topological type of the original bundle.
From the f\/ield-theoretical point of view the modif\/ication is
provided by a
 gauge transformation of sections, which is singular at the divisor.
 In~\cite{LOZ1} we apply this procedure
to the Higgs bundles. The Higgs bundles are the phase spaces of the
Hitchin integrable systems~\cite{Hi}. Modif\/ications acts on the
phase space as a symplectic transformation. In this special case we
call the modif\/ication the Symplectic Hecke Correspondence. For the
Higgs bundles over elliptic curves with marked points Symplectic
Hecke Correspondence leads to
 a symplectomorphism between dif\/ferent classical integrable systems
such as
\begin{itemize}\itemsep=0pt
\item Elliptic Calogero--Moser system $\Leftrightarrow$
Elliptic $\GLN$ Top,~\cite{LOZ1};
\item Calogero--Moser f\/ield theory $\Leftrightarrow$
Landau--Lifshitz equation,~\cite{LOZ1,Kr1};
\item Painlev\'{e} VI $\Leftrightarrow$ non-autonomous
Zhukovsky--Volterra gyrostat,~\cite{LOZ2}.
\end{itemize}

In these examples modif\/ications increase the degree of the
underlying bundles on one. In general,
 modif\/ications act as the B\"{a}cklund transformations of integrable systems.
If degree of the bundles (modula rank) is not changed then
modif\/ications produce what is called the autoB\"{a}cklund
transformations.
 It turned out that the modif\/ication in the f\/irst example is equivalent to
 the twist of $R$-matrices \cite{B,JMO}
  that transforms the dynamical $R$-matrices of the
 IRF models of the ${\rm GL}(N)$ type \cite{Fe} to the vertex $R$-matrices~\cite{Be} corresponding to the ${\rm GL}(N)$ generalization of the XYZ models.

The modif\/ications are parameterized by vectors $\vec m$ of the
weight lattices $P$ of $\SLN$. If~$\vec m$ belongs to the root
sublattice $Q\subset P$, then the modif\/ied bundle $\ti E$
 has the same degree as $E$. Otherwise, the degree of bundle
 is changed. The modif\/ications can be described by changing another topological invariant.
 It is a
 characteristic class of a bundle. Let the base of $E$  be a Riemann surface $\Si_g$ of genus $g$.
Then the characteristic class of $E$  is an element of
$H^2(\Si_g,\mZ_N)\sim\mZ_N$, where $\mZ_N\sim P/Q$ is a center of
$\SLN$. Another example of  the characteristic classes, is the
characteristic class of spin-bundles, that will not considered here,
is the Stiefel--Whitney class $H^2(\Si_g,Z_2)$.

Here we discuss a f\/ield-theoretical interpretation of modif\/ications.
It was established  in~\cite{KW} that the modif\/ications are
related to the
 Dirac monopole conf\/igurations in a topological version of the $\clN=4$ four-dimensional
super-symmetric Yang--Mills theory.
If ``the space-time'' of the topological theory has the form
 $\mR^2\times\Si_g$,
  then  the modif\/ications of $E$ over  $\Si_g$ are parameterized
 by  the  monopoles charges.

To describe the modif\/ication it is suf\/f\/icient to neglect the ``time''
dependence and consider $\mR\times\Si_g$. The condition for f\/ields
to preserve the supersymmetry amounts to the Bogomolny equation.

The aims of this paper are
\begin{itemize}\itemsep=0pt
  \item To def\/ine modif\/ications and describe their interrelations with
  the Bogomolny equation following~\cite{KW}.
  We consider a special conf\/iguration of the
space-time $\mR^2\times\Si_\tau$, where $\Si_\tau$ is an elliptic
curve with the modular parameter $\tau$.

  \item To f\/ind solutions of the Bogomolny equation in the case of
  line bundles over $\Si_\tau$. They are
  generalizations of the Kronecker series \cite{We}.
We give two representations of the solution and prove their
equivalence by means of the functional equation generalizing the
Kronecker functional equation.

  \item To describe non-Abelian modif\/ications that are not related directly to
solutions of the Bogomolny equation and follows from our previous
results.
\end{itemize}

\section{Characteristic classes of holomorphic bundles\\ over complex curves}\label{section2}

We describe holomorphic bundles over  complex curves  $\Si_g$ of
genus $g$ and def\/ine their characteristic classes.

\subsection{Global description}\label{section2.1}

Let  $\pi_1(\Si_g)$ be a fundamental group of $\Si_g$.
It has
 $2g$ generators $\{a_\al,b_\al\}\,$, corresponding to the
 fundamental cycles of $\Si_g$ with the relation
\begin{gather}\label{0.3}
\prod_{\al=1}^g [a_{\al},b_{\al}]=1,
\end{gather}
where
$[a_{\al},b_{\al}]= a_{\al}b_{\al}a_{\al}^{-1}b_{\al}^{-1}$ is the
group commutator.

Let $\rho$ be a representation of $\pi_1$ in $\mC^N$.
Consider a holomorphic adjoint $\GLN$ bundle $E$ over $\Si_g$. In
fact, $E$ is a P$\GLN\sim\PSLN$ bundle, because the center of $\GLN$
does not act in the adjoint representation. The bundle $E$ can be
def\/ined by holomorphic transition matrices of its sections
$s\in\G(E)$ around the fundamental cycles. Let $z\in \Si_g$ be a
f\/ixed point. Then
\begin{gather*}
 s(a_\al z)= \rho(a_\al)s(z),\qquad
 s(b_\be z)= \rho(b_\be)s(z).
\end{gather*}

Due to (\ref{0.3}) we have \begin{gather}\label{pi} \prod_{\al=1}^g
[\rho(a_{\al}),\rho(b_{\al})]={\rm Id}.
\end{gather}

Let $\clK$ be an extension of  $\pi_1$ by the  cyclic group
$\mZ_N\sim\mZ/N\mZ$
\begin{gather}\label{eseq} 1\to\mZ_N\to\clK\to\pi_1(\Si_g)\to
1. \end{gather}
The group $\clK$ is def\/ined by the relation
\begin{gather*}
\prod_{\al=1}^g [a_{\al},b_{\al}]=\om,\qquad \om^N=1.
\end{gather*}
Let
$\hat{\rho}$  be a representation of $\clK$ in $\GLN$. Then using
$\hat{\rho}$ as transition matrices
 we def\/ine a~bundle over $\Si_g$. But now
instead of (\ref{pi}) we have \begin{gather}\label{pi1} \prod_{\al=1}^g
[\hat\rho(a_{\al}),\hat\rho(b_{\al})]=\om\, {\rm Id}. \end{gather} Here $\om\,
{\rm Id}$ is the generator of the center $\clZ(\SLN)\sim\mZ_N$ of $\SLN$.
It means that $\hat\rho$ can serve as transition matrices only for
 $\PSLN$ bundles,
but not for $\SLN$ or $\GLN$ bundles. Note, that the f\/ibers of the
$\PSLN$-bundles are spaces of representations with highest weights
from the root lattice $Q$ (\ref{rl}) including the adjoint
representation with the highest weight $\varpi_1+\varpi_{N-1}$
(\ref{fw1}). For the $\SLN$ representations the highest  weights
belong to the weight lattice $P$ (\ref{wl}).
 In this way elements from  the factor group $P/Q\sim \clZ(\SLN)$ (\ref{center}) def\/ine
 an obstruction to lift  $\PSLN$ bundles to $\SLN$ bundles.

The obstruction has a cohomological interpretation.
 Consider the exact sequence  following from (\ref{a1})
 \begin{gather*}
\to H^1(\Si_g,\SLN)\to H^1(\Si_g,\PSLN)\to
H^2(\Si_g,\clZ(\SLN))\to\cdots.
 \end{gather*} The groups $H^1(\Si_g,\SLN)$,
$H^1(\Si_g,\PSLN)$
are the moduli space of $\SLN$ and
 \linebreak $\PSLN$ bund\-les. Then $H^2(\Si_g,\clZ(\SLN))$
def\/ines an obstruction to lift  $\PSLN$ bund\-les to $\SLN$ bund\-les.
We call $\xi\in H^2(\Si_g,\mZ_N)$ \emph{the characteristic class} of
a $\PSLN$ bundle. In fact, $H^2(\Si_g,\mZ_N)\sim\mZ_N$ and $\om$ in
(\ref{pi1}) represents $\xi\in H^2(\Si_g,\mZ_N)$.

This construction can be generalized to any factor-group
$G_l=\SLN/\mZ_l$, where $l$ is a~nontrivial divisor of $N$,
$(N=pl$, $l\neq 1,N)$. Consider an extension $\clK_l$ of
$\pi_1(\Si_g)$ by $\mZ_l$ (compare with (\ref{eseq}))
\begin{gather*}
1\to\mZ_l\to\clK_l\to\pi_1(\Si_g)\to 1.
\end{gather*}
 Let $E_l$ be a holomorphic $G_l$-bundle. The f\/ibers of~$E_l$ belong to
 a irreducible representation of~$G_l$ with a highest weight $\nu\in\G(G_l)$
 (\ref{gl}).
Then the transition matrices representing $\clK_l$ satisfy the
relation \begin{gather}\label{pi2} \prod_{\al=1}^g
[\hat\rho(a_{\al}),\hat\rho(b_{\al})]=\om^p\, {\rm Id},\qquad (\om^p)^l=1.
\end{gather} It follows from the exact sequence \begin{gather*}
1\to\mZ_l\to\SLN\to G_l\to 1, \end{gather*} that elements from
$H^2(\Si_g,\mZ_l)\sim\mZ_l$ are obstructions to lift $G_l$ bundle
$E_l$ to a $\SLN$-bundle. The group $\mZ_l$ can be identif\/ied with
the center of the dual group $^LG_l\sim G_p=\SLN/\mZ_p$ (see
(\ref{dugr}) and (\ref{cdg})). Thus, the obstructions to lift $G_l$
bundles $E_l$ to a $\SLN$ bundles are def\/ined by
$H^2(\Si_g,\clZ(^LG_l))$.

On the other hand, since  $\mZ_p$ is a center of $G_l$ we have the
sequence
\begin{gather*}
1\to\mZ_p\to G_l\to\PSLN\to 1,
 \end{gather*}
where $\mZ_p$ is a center of $G_l$. Then elements from
$H^2(\Si_g,\clZ(G_l))$ are obstructions to lift a $\PSLN$-bundle to
a $G_l$-bundle. Summarizing we have def\/ined two types of the
characteristic classes
\begin{gather}
H^2(\Si_g,\clZ(G_l)) - {\rm
obstructions~to~lift~a~}\PSLN~{\rm bundle~to~a~}G_l~{\rm bundle},\nonumber\\
\label{obs2}
H^2(\Si_g,\clZ(^LG_l)) - {\rm obstructions~to~lift~a~}G_l~{\rm bundle~to~a~}\SLN~{\rm bundle}.
\end{gather}

Though for $\om\neq 1$   $\PSLN$ bundles cannot be lifted to $\SLN$
bundles, they can be lifted to $\GLN$ bundles. From the exact
sequence
\begin{gather*}
1\to \clO^*\stackrel{\mathrm{det}}\to\GLN\to {\rm
PGL}(N,\mC)\to 1
\end{gather*}
we have
\begin{gather*}
H^1(\Si_g,\GLN)\to
H^1(\Si_g,{\rm PGL}(N,\mC))\to H^2(\Si_g,\clO^*).
\end{gather*} The Brauer
group $H^2(\Si_g,\clO^*)$ vanishes and therefore, there is no
obstruction to lift ${\rm PGL}(N,\mC)\sim{\rm PSL}(N,\mC)$ bundles
to $\GLN$ bundles. We will demonstrate it below.

\subsection{Holomorphic bundles over elliptic curves}\label{section2.2}

We def\/ine an elliptic curve ($g=1$) as the quotient
$\Si_\tau=\mC/(\mZ+\tau\mZ)$. In this case we can construct
explicitly the generic transition matrices for $G_l$-bundles.

The curve has two fundamental cycles $a: (z\to z+1)$,
$b: (z\to z+\tau)$. We def\/ine a trivial bundle  $E$ over
$\Si_\tau$ by two commuting matrices
\begin{gather}\label{14}
s(z+1)=\rho_as(z),\qquad s(z+\tau)=\rho_bs(z),\qquad [\rho_a,\rho_b]={\rm Id}.
\end{gather} It is a $\PGLN$-bundle that can be lifted to $\SLN$ bundles.

Consider a representation of $\hat\rho$ of $\clK$ acting on the
sections of $E$ as \begin{gather*}
s(z+1)=\hat{\rho}_as(z),\qquad s(z+\tau)=\hat{\rho}_bs(z).
\end{gather*} with
commutation relation  (\ref{pi1})
\begin{gather*}
[\hat{\rho}_a,\hat{\rho}_b]=\om\,{\rm Id}. \end{gather*}
One can choose
\begin{gather}\label{qla}
\hat{\rho}_a=\clQ,\qquad \hat{\rho}_b=\La, \qquad
\clQ=\di\big(1,\om,\ldots,\om^{N-1}\big),\qquad
\La=\left(
      \begin{array}{cccc}
        0 & 1 & \ldots & 0 \\
        0 & 0 & 1 & 0 \\
        \vdots & \vdots &\ddots  & 1 \\
        1 & 0 & \ldots & 0 \\
      \end{array}
    \right).
\end{gather} The bundle with these transition functions cannot be lifted to
$\SLN$ bundles. Replace $\hat{\rho}_b$ by \begin{gather}\label{last}
\hat{\rho}'_b=\exp\left(-\frac{2\pi
i}{N}\left(z+\frac{\tau}{2}\right)\right)\La.
\end{gather} It is a $\GLN$ bundle since
$[\hat{\rho}_a,\hat{\rho}'_b]=Id$ and $\det\hat{\rho}'_b\neq 1$ . It
follows from (\ref{last}) that a section of the determinant bundle
is the theta-function \begin{gather}\label{theta} \vartheta(z,\tau)=q^{\frac
{1}{8}}\sum_{n\in {\mZ}}(-1)^ne^{\pi i(n(n+1)\tau+2nz)},\qquad q=\exp
2\pi i\tau. \end{gather} It has a simple pole in the fundamental domain
$\mC/(\mZ\oplus\tau\mZ)$. Therefore, the bundle has  degree one. It
is called the theta-bundle.

To consider a general case \cite{LZ} represent the rank as the
product $N=pl$. Def\/ine the transition matrix
\begin{gather}\label{ic}
\hat\rho_a=\clQ , \\
\label{ic1} \hat\rho_b=\bfe(\vec u_l)\La^p,
\end{gather}
 where \begin{gather*}
 \vec u_l=\di (\overbrace{\bfu_p,\ldots,
\bfu_p}^l),\qquad \bfu_p=(\ti u_1,\ldots,\ti u_p). \end{gather*} Since
$[\clQ,\La^p]=\om^p\,{\rm Id}_l$, $\om^p=\exp\frac{2\pi i}{l}$
\begin{gather*}
[\hat\rho_a,\hat\rho_b]=\om^p\,{\rm Id}_N.
\end{gather*} Comparing this
relation with (\ref{pi2}) we conclude that (\ref{ic}) and
(\ref{ic1}) serve as the transition matrices for a $G_l$-bundle over
$\Si_\tau$. Therefore $\om^p$ represents an element from
\mbox{$H^2(\Si_\tau,\clZ(^LG_l))\sim \mZ_p$}. It is an obstruction
(\ref{obs2}).

As in (\ref{last}), modify   the transition matrix \begin{gather*}
\hat\rho_b\to\hat\rho_b'=\exp\left\{-\frac{2\pi
i}{p}\left(z+\frac{\tau}{2}\right)\right\} \hat\rho_b. \end{gather*} We come to the
$\GLN$-bundle of  degree $p$ (mod$\,N$).

\subsection{Local description}\label{section2.3}

There exists another description of a holomorphic bundles over
$\Si_g$. Let $w_0$ be a f\/ixed point on $\Si_g$
 and $D_{w_0}$ ($D^\times_{w_0}$) be a disc (punctured disc) with a center $w_0$
with a local coordinate $z$. A~bundle~$E$ over~$\Si_g$ can be
trivialized over  $D$ and over $\Si_g\setminus w_0$. These two
trivializations are related by a $\GLN$ transformation $g(z)$,
holomorphic on $D^\times_{w_0}$. If we consider another
trivialization over $D$ then $g$ is multiplied from left by an
invertible matrix $h$ on  $D$. Likewise, a~trivialization over
$\Si_g\setminus w_0$ is determined up to the multiplication on the
right $g\to gh$ , where $h\in\GLN$ is holomorphic on
 $\Si_g\setminus w_0$.
Thus, the set of isomorphism classes of rank $N$ vector bundles is
described as a double-coset{\samepage
\begin{gather*}
\GLN(D_{w_0})\setminus
\GLN(D^\times_{w_0}) /\GLN(\Si_g\setminus w_0), \end{gather*} where
$\GLN(U)$ denote the group of $\GLN$-valued holomorphic functions on
$U$.}

 Let $\det\,g(z)=1$.
 If $g(ze^{2\pi i})=g(z)$ then it def\/ines a $\SLN$-bundle over $\Si_g$.
 But if the monodromy is nontrivial
\begin{gather}\label{abc11}
 g(ze^{2\pi i})=\om g(z),\qquad \om^N=1,
 \end{gather}
 then $g(w)$ is a transition matrix for a $\PSLN$-bundle but not for a $\SLN$-bundle.
  This relation is similar to (\ref{pi1}).

Let us choose a trivialization of $E$ over $D$ by choosing $N$
linear independent holomorphic sections $\vec
s=(s_1,s_2,\ldots,s_N)$. Thereby, the bundle $E$ over $D$ is
represented by a sum of $N$ line bundles
$\clL_1\oplus\clL_2\oplus\cdots\oplus\clL_N$. The sections over
 $\Si_g\setminus w_0$ are obtained by the action of the
 transition matrix $\vec s\,'=\vec s g$.

Let $\vec m$ belongs to the root lattice $(\vec
m=(m_1,m_2,\ldots,m_N)\in Q)$ (\ref{rl}). Transform the restriction
of the section $\vec s$ on  $D^\times_{w_0}$ as
 \begin{gather}\label{mod}
 s_j\to z^{-m_j}s_j, \qquad j=1,\ldots,N.
\end{gather} Then the transition matrix is transformed by the diagonal matrix
\begin{gather}\label{mtm}
g(z)\to \di\big(z^{-m_1},z^{-m_2},\ldots,z^{-m_N}\big)g(z).
\end{gather}
It implies the transformation of line bundles over $D$ \begin{gather*}
\clL_j\to \clL_j\otimes \clO(m_j). \end{gather*} In this way we come to the
new bundle $\ti E$ (\emph{the modif\/ied bundle}). It is def\/ined by
the new transition matrix~(\ref{mtm}).
This transformation of the bundle $E$ to $\ti E$ (or more exactly
the map of sheaves of its sections)
\begin{gather*}
\G(E)\,\mapright{\Xi(\vec m)} \,\G(\ti E),\qquad  \Xi(\vec
m)\sim\di\big(z^{-m_1},z^{-m_2},\ldots,z^{-m_N}\big),
 \end{gather*}
 is called \emph{the modification
or the Hecke transformation} of type $\vec m=(m_1,m_2,\ldots,m_N)$.
In f\/ield-theoretical terms it corresponds to \emph{the t'Hooft
operator,} generating by monopoles (see below).

Let us relax the condition $\sum_{j=1}^N m_j=0$. Then the
modif\/ication $\Xi(\vec m)$ changes the topology of $E$. We come to a
nontrivial bundle of degree ${\rm deg}\,(\ti E)={\rm deg}\,(E)+\sum_{j=1}^N
m_j$. In next section we illustrate this fact.

Now assume that $\vec m$ belongs to the weight lattice $P$
(\ref{wl}). Then the modif\/ication $\Xi(\vec m)$ changes the
characteristic class of a $\PGLN$-bundle $E$. To prove it let us
pass to the basis of the fundamental weights (\ref{fw1})
\begin{gather*}
\vec
m=\sum_{j=1}^{N}m_je_j=\sum_{k=1}^{N-1}n_k\varpi_k. \end{gather*}

 It follows from (\ref{fw1}) that $m_j$ and $n_k$ are related as
\begin{gather*}
  m_1=\f1{N}((N-1)n_1+(N-2)n_2+\dots+n_{N-1}),\nonumber \\
  m_2=\f1{N}(-n_1+(N-2)n_2+\dots+n_{N-1}),\nonumber \\
 \cdots\cdots\cdots\cdots\cdots\cdots\cdots  \cdots\cdots\cdots\cdots\cdots\cdots\cdots \nonumber\\
  m_N=\f1{N}(-n_1-2n_2-\dots-(N-1)n_{N-1}),
\qquad n_k=m_k-m_{k+1}. 
\end{gather*}
 Rewrite the modif\/ication in the
form of the product of the diagonal matrices \begin{gather}\label{gtr} \Xi(\vec
n)\sim \prod_{k=1}^{N-1}\di\big(z^{-n_k\varpi_k}\big). \end{gather} It follows
from (\ref{fw1}) that the monodromy of this matrix around the point
$z=0$ is \begin{gather}\label{mo1} \exp\left(-\frac{2\pi
i}{N}\sum_{k=1}^{N-1}kn_k\right){\rm Id}_N.
\end{gather} Therefore, the
characteristic class of the adjoint bundle is unchanged if
\begin{gather*}
\sum_{k=1}^{N-1}kn_k=N\sum_{j=1}^{N-1}m_j=0,
\qquad ({\rm mod}\,N).
\end{gather*}
In this case the weight vector $\vec m$
 belongs to the root lattice $Q$. Otherwise, we come to the non-trivial monodromy
 (\ref{mo1}). It is an obstruction to lift the $\PGLN$-bundle to a $\SLN$-bundle.
  This element can be identif\/ied with the monodromy (\ref{pi1}) and
in  this way with an element from $H^2(\Si,\mZ_N)$.
 As it was mentioned above, the modif\/ied bundle
$\ti E$ can be lifted to a~$\GLN$ bundle. Let us act on the modif\/ied
sections (\ref{mod}) by the scalar matrix \begin{gather*}
h=z^{\frac{2\pi i}{N}\sum\limits_{k=1}^{N-1}kn_k}\,{\rm Id}_N.
\end{gather*} It is a
$\GLN$ gauge transformation. The monodromy of the new transition
matrix is trivial. Therefore, we come to the $\GLN$ bundle. The
bundle is topologically nontrivial -- it has degree
\begin{gather}\label{edgr}
p=\sum_{k=1}^{N-1}kn_k=N\sum_{j=1}^{N-1}m_j.
\end{gather}
 It follows from (\ref{mo1}) that
the characteristic class $\xi$ and the degree $p$ are related as
\begin{gather*}
\xi=\exp\frac{2\pi i}{N}p.
\end{gather*} The set of
modif\/ications that changes the degree on $p$ is def\/ined as solutions
of (\ref{edgr})
 in integers~$n_k$.

Assume that the bundle $E$ is equipped with a holomorphic
connection. On $D^\times_{w_0}$ it takes the form $(\p_z+A_z) dz$
and can be considered as an element of the af\/f\/ine Lie coalgebra
$\widehat{\rm gl}^*(N,\mC)(D^\times_{w_0})$ The gauge transformation
(\ref{mtm}) acts on $A_w$ acts as the coadjoint action \begin{gather}\label{a}
(A_z)_{jk}dz \to \big(z^{m_k-m_j}(A_z)_{jk}(1-\de_{jk})-m_jz^{-1}\de_{jk}\big)dz.
\end{gather} Let $\vec m\in P$. Then the f\/irst term in the r.h.s.\ is well
def\/ined, since $m_k-m_j$ is integer. The last term represents the
shift action~(\ref{shi}) of the af\/f\/ine group $\bar W_a$ (\ref{q25})
on the connection. The topology of $E$ is not changed if $\vec m\in
Q$ and we come to description of the characteristic class as
elements from factor group $\bar W_a/W_a$ (\ref{fwe}). We come again
to this point in Section~\ref{section4}.

Let $N=pl$ with $l\neq 1,N$ and $G_l=\SLN/\mZ_l$ (\ref{gg}).
Consider the gauge transformation~(\ref{gtr}) with $\vec m$
$(\vec\varpi)\in\G(^LG)$~(\ref{gg}). For example, we can take
$\vec\varpi=(p,0,\ldots,0)$. Then the monodromy (\ref{mo1}) belongs
to the group $\mZ_l$. It means that the modif\/ied bundle $\ti E$ is
the $G_l$-bundle that cannot be lifted to the $\SLN$-bundle (see~(\ref{obs2})).

The modif\/ication can be performed in an arbitrary number of points
$w_{a}$, $(a=1,\ldots,n)$. To this end def\/ine the isomorphism
classes of vector bundles as the quotient \begin{gather*}
\prod_{a=1}^n\GLN(D_{w_a}) \setminus
\prod_{a=1}^n\GLN(D^\times_{w_a}) /\GLN(\Si_g\setminus
(w_1,\ldots,w_a)). \end{gather*} We have $n$ transition matrices $g_a(z_a)$
representing an element of the quotient, where $z_a$ is a local
coordinate. Let $\Xi(\vec m_a)$ denotes the modif\/ication of $E$ at
$w_{a}$ and $\Xi=\prod_{a=1}^n\Xi(\vec m_a)$. The order of
modif\/ications in the product is irrelevant, since they commute. To
calculate the monodromy of $\Xi$ we choose the same orientation in
all points $w_a$. The characteristic class of $\xi$ of modif\/ied
bundle $\ti E$ corresponds to \begin{gather*}
\prod_{a=1}^n\exp\left(-\frac{2\pi
i}{N}\sum_{k=1}^{N-1}kn_k^a\right). \end{gather*}

\section{Bogomolny equation}\label{section3}

\textbf{Def\/inition.}
Let $W=\mR\times\Si_g$. Consider a bundle $V$ over $W$ equipped with
the curvature $F$. Let $\phi$ be a zero form on $W$ taking value in
sections of the adjoint bundle $\phi\in\Om^0(W,{\rm End}\, V)$. It is the
so-called Higgs f\/ield.

 The Bogomolny equation on $W$
takes the form \begin{gather}\label{q1} F=*D\phi. \end{gather}
Here $*$ is the Hodge
operator on $W$  with respect to the metric $ds^2$ on $W$. In local
coordinates $(z, \bz)$ on $\Sigma_g$ and $y$ on the real line
 $ds^2=g|dz|^2+dy^2$,
where $g(z,\bz)|dz|^2$ is a metric on $\Si_g$. Then the Hodge
operator is def\/ined as \begin{gather*}
\star dy=\tfrac 12 i gdz\wedge d\bz,\qquad
\star dz= -i dz\wedge dy, \qquad \star d\bz= i d\bz\wedge dy,
\end{gather*} and
(\ref{q1}) becomes
\begin{subequations}\label{q2}
\begin{gather}\label{q2-1}
\p_{z} A_{\bz}-\p_{\bz} A_z+[A_z,A_{\bz}]=\frac{ig(z,\bz)}{2}\left(\p_y\phi+[A_y,\phi]\right),
\\
\label{q2-2}  \p_y A_z-\p_{z} A_y+[A_y,A_z]=i(\p_{z}\phi+[A_z,\phi]),
\\
\label{q2-3}  \p_yA_{\bz}-\p_{\bz}
A_y+[A_y,A_{\bz}]=-i(\p_{\bz}\phi+[A_{\bz},\phi]).
\end{gather}
\end{subequations} In what follows we will consider only $\PSLN$-bundles.

A monopole solution of this equation is def\/ined in the following way.
 Let $\ti{W}=(W\setminus \vec{x}^0=(y=0,z=z_0))$.
The Bianchi identity $ DF=0$ on $\ti{W}$ implies that $\phi$ can be
identif\/ied with the Green function for the operator $\star D\star D$
\begin{gather}\label{GF}
 \star D\star D\phi=M\de(\vec{x}-\vec{x}^0),\\
 \label{mm}
M=\di(m_1,m_2,\ldots,m_N)\in\gln,\qquad \vec m=(m_1,m_2,\ldots,m_N)\in
P \quad \eqref{wl},
\end{gather} and $(m_1,m_2,\ldots,m_N)$ are the monopole
charges.
 We explain below this choice of $M$.
This equation means that $\phi$ is singular at $\vec{x}^0$.

\textbf{Boundary conditions and gauge symmetry.} In what follows except Section~\ref{section3.1}
 we assume that  $\p_y\phi$ vanishes when $y\to \pm\infty$.
It is the Neumann boundary conditions for the Higgs f\/ield, while the
gauge f\/ields are unspecif\/ied.
 Let $V_\pm$ be restrictions of $V$ to the bundles over $\Si_g$
 on the ``left end'' and ``right end'' of
$W: y\to \pm\infty$. These bundles are f\/lat.
 It follows from~(\ref{q5-1}), where the gauge $A_y=0$ is assumed.
  It was proved in \cite{KW} that
  in  absence of  the source $M=0$ in (\ref{GF}) the only solutions of~(\ref{q1})
with these boundary conditions  are $F=0$, $\phi=0$. Note that
these boundary conditions dif\/fer from ones chosen in~\cite{KW}.

The Bogomolny equation def\/ines a transformation $V_-\to V_+$. ($E$
and $\ti E$ in our notations in Introduction.) We will see in next
sections that in general the characteristic classes of bundles are
changed under these transformations. It depends on the monopole
charges $\vec m=(m_1,m_2,\ldots,m_N)$.

The system (\ref{q2}) is invariant with respect to the gauge group
$\clG$ action:
\begin{gather}
A_z\rightarrow hA_zh^{-1}+\p_{z} hh^{-1} , \qquad A_{\bz}\rightarrow
hA_{\bz} h^{-1}+\p_{\bz} hh^{-1}, \nonumber\\
 A_y\rightarrow hA_yh^{-1}+\p_y hh^{-1}, \qquad
\phi\rightarrow h\phi h^{-1},\label{q3}
\end{gather} where $h\in\clG$ is a smooth map $W\to\GLN$. To preserve the
r.h.s.\ in (\ref{GF}) it should satisfy the condition
$[h(\vec{x}^0),M]=0$.

Assume for simplicity that $V$ is an adjoint bundle. Since the gauge
f\/ields for $y=\pm\infty$ are unspecif\/ied and only f\/lat we can act on
them by boundary values of the gauge group
\mbox{$\clG|_{y=\pm\infty}=\clG_\pm$}. Then $\clM_\pm=\{V_\pm\}/\clG_\pm$
are the moduli spaces of f\/lat bundles.

\textbf{Relations to integrable systems.}
The moduli spaces of f\/lat bundles are phase spaces of non-autonomous
Hamiltonian systems related to the isomonodromy problems over
 $\Si_g$. The isomonodromy problem takes the form
 \begin{gather}\label{ip}
   [\p_{z}+A_z,\Psi]=0    ,\qquad
  [\partial_{\bz}+A_{\bz},\Psi]=0.
 \end{gather}
 Here $\Psi\in\Om^0(\Si_g,{\rm Aut}\, V)$ is the Baker--Akhiezer function. These system is compatible
 for any degree of bundle, because it is def\/ined in the adjoint representation.
 One example of these systems
 we have mentioned in Introduction
($V_-\to$ Painlev\'{e} VI) and ($V_+\to$ Zhukovsky--Volterra
gyrostat).

It is known, that the moduli space of f\/lat bundles are deformation
(the Whitham deformation) of the phase spaces of the Hitchin
integrable systems~-- the moduli spaces of the Higgs bundles. To
consider this limit one should replace a holomorphic connection by
the $\kappa$-connection $\kappa\p_{z}+A_z$
 introduced by P.~Deligne and take a limit $\kappa\to 0$.
 It is a quasi-classical limit in the linear problem~(\ref{ip}).
 Details can be found in~\cite{FN,Kr,LO,Ar}.
 In this way a monopole solution put in a correspondence (symplectic Hecke correspondence)
 two Hitchin systems
(the f\/irst and the last examples in Introduction). But Bogomolny
equation tells us more. It describes an evolution from one type of
system to another.

It is possible to generalize (\ref{GF}) and consider multi-monopole
sources $\sum_aM_a\de(\vec{x}-\vec{x}_a^0)$  in the r.h.s. This
generalization will correspond to modif\/ications in a few points of
$\Si_g$ described at the end of previous section.

It is interesting that in some particular cases this situation was
discussed in the frameworks of a supersymmetric Yang--Mills theory~\cite{SY,Po}\footnote{We are grateful to A.~Gorsky who bring our
attention to this point.}. It was observed there that a monopole
conf\/iguration corresponds to a soliton type evolution along $y$.
Therefore, it can be suggested that the system (\ref{q2}) is
integrable. We did not succeed to prove this fact, but propose a
linear problem related to the Bogomolny equation. An associated
linear problem allows one in principal to apply the methods of the
Inverse Scattering Problem or the Whitham approximation to f\/ind
solutions \cite{ZMNP}. Assume that the metric $g$ on $\Si_g$ is a
constant. Then   the system (\ref{q2}) is
 the compatibility condition for the linear system
\begin{gather*}
\left(\p_z+A_z+\tfrac 12 \la^{-1}g(\p_y+A_y+i\phi)\right)\psi=0,\nonumber\\
\left(\p_{\bz}+A_{\bz}+\tfrac 12 \la g(\p_y+A_y-i\phi)\right)\psi=0,
\end{gather*} where $\la\in\mC P^1$ is a spectral parameter.
 It can be suggested that monopole solution of (\ref{q2})
corresponds to a soliton solution of this system. We will not
develop here this approach\footnote{The SU(2) case and $W=\mR^3$
was analyzed in~\cite{Wa} for dif\/ferent boundary conditions.}.

\textbf{Gauge f\/ixing.}
Choose a gauge f\/ixing conditions as: $A_{\bz}=0$. Holomorphic
functions $h=h(y,z)$ preserve this gauge. Then
\begin{gather*}
-\p_{\bz} A_z=\frac{ig}{2}\left(\p_y\phi+[A_y,\phi]\right),\nonumber
\\
\p_y A_z-\p_{z} A_y+[A_z,A_y]=i(\p_{z}\phi+[A_z,\phi]),\nonumber
\\
\p_{\bz} A_y=i\p_{\bz}\phi.
\end{gather*} The last equation means that $A_y-i\phi$ is holomorphic.
It follows from (\ref{q3}) that the gauge transformation of this
function is
\begin{gather*}
A_y-i\phi\rightarrow h(A_y-i\phi)h^{-1}+\p_y
hh^{-1}.
 \end{gather*} Thus, we can keep $A_y=i\phi$ by using holomorphic and
$y$-independent part of the gauge group $(\p_yh=0)$. Finally, we
come to the system
\begin{subequations} \label{q5}
\begin{gather}
\label{q5-1}  \p_{\bz} A_z=-\frac{ig}{2}\p_y\phi,
\\
\label{q5-2}  \p_y A_z-2i\p_{z} \phi+2i[A_z,\phi]=0,
\\
\label{q5-3}  A_y=i\phi,
\\
\label{q5-4}  A_{\bz}=0.
\end{gather}
\end{subequations} Two upper equations from (\ref{q5}) lead to the Laplace
type equation \begin{gather}\label{q6}
\p^2_y\phi+\frac{4}{g}(\p_{z}\p_{\bz}\phi+\p_{\bz}[A_z,\phi])=0.
\end{gather}

In scalar case (\ref{q6}) is simplif\/ied \begin{gather}\label{q7}
\p^2_y\phi+\frac{4}{g}\p_{z}\p_{\bz}\phi=0. \end{gather}

\subsection{Rational solution in scalar case}\label{section3.1}

In this subsection we replace $\Si_g$ by $\mC$. The coordinates
$z,\bz$ on $\mC$ will play the role of local coordinates on $\Si_g$.
 Consider (\ref{q7}) on
$\ti W=\mR\times\mC\setminus(0,0,0)$. In this particular case we can
choose the boundary conditions in the following form:
\begin{gather}\label{q8}
\phi|_{y=\pm\infty}=0, \\
\label{q9}
 A_z|_{y=\pm\infty}=0.
\end{gather} The solution of (\ref{q7}) with $g=1$ satisfying (\ref{q8})
 has the form:
\begin{gather}\label{q10} \phi=c\frac{1}{\sqrt{y^2+z\bz}}, \end{gather} where $c$ is a
constant. So in fact we deal here with the Laplace equation on
$\mathbb{R}\times\mathbb{C}\setminus(0,0,0)$. It follows from
(\ref{q9}) and from the equation $\p_{\bz} A_z=-\frac{i}{2}\p_y\phi$
(\ref{q5-1}) that
\begin{gather}
A_z(z,\bar{z},y)=A_z^+(z,\bar{z},y),\qquad  y>0\quad \mbox{and}\quad y=0, \quad z\neq 0,\nonumber
\\
A_z(z,\bar{z},y)=A_z^-(z,\bar{z},y),\qquad y<0,\label{q101}
\end{gather}
 where \begin{gather*}
A_z^+(z,\bar{z},y)=-ic\left(\frac{1}{z}\frac{y}{\sqrt{y^2+z\bz}}-\frac{1}{z}\right)
+{\rm const},\nonumber
\\
A_z^-(z,\bar{z},y)=-ic\left(\frac{1}{z}\frac{y}{\sqrt{y^2+z\bz}}+\frac{1}{z}\right)
+{\rm const},
\end{gather*}
and $A_z(z,\bar{z},y)$ is a connection on the line bundle $\clL$
over $\ti W$. The connection has a jump $-2ic\frac{1}{z}$ at $y=0$.
To deal with smooth connections we compensate it by a holomorphic
gauge transform that locally near $\vec x_0$ has the form $h\sim
z^m$. Here $m$ should be integer, because $h$ is a smooth function.
Notice that all holomorphic line bundles over  $S^2$ are known to be
 $\mathcal{O}(m)$-bundles, $m\in\mZ$.
Thus, we have $c=i\frac{m}{2}$,  $m\in\mathbb{Z}$. This usually
referred as a quantization of the monopole charge. In fact the
constant $c$ contains factor $4\pi$ (area of a unit sphere) which
yields a proper normalization of delta-function and appears in
Gauss's law. The gauge transformation $h$ is the modif\/ication
(\ref{mod}), (\ref{mtm}) for line bundles over $\mC P^1$. This is
what we mean saying that the described 3-dimensional construction
characterizes the modif\/ication of the corresponding bundle.

Consider for a moment the general situation $ W=\mR\times\Si_g$ and
let $z,\bz$ be local coordinates on~$\Si_g$. Locally near $\vec
x_0=(0,0,0)$ connections corresponding to
 solutions of~(\ref{q7}) have the form~(\ref{q101}).
 Let $S^2$ be a small sphere surrounding the point $\vec x_0$ in $W$ and
 $\Si_{g,\pm}$ be the left and
right boundaries of~$W$ and~$\clL_{\pm}$ are the corresponding
restrictions of $\clL$. Then (as it is explained in \cite{KW} in
detail)
\begin{gather*}
\int_{\Si_{g,+}}F=\int_{\Si_{g,-}}F +m,
\end{gather*}
where $F$ is a curvature of the connection $A$. In other words, the
monopole solution with the charge $m$ increases the degree of bundle
by $m$ $({\rm deg}\, \clL_+={\rm deg}\, \clL_-+m)$.

\subsection{Elliptic solution in scalar case}\label{section3.2}

 The Laplace equation (\ref{q7}) on $\Si_\tau$ has the form
\begin{gather}\label{q111} \p^2_y\phi+4({\mathcal
Im}(\tau))^2\p_{z}\p_{\bz}\phi=0,
\end{gather} or
\begin{gather*}
\p^2_y\phi+(2\pi\al)^2\p_{z}\p_{\bz}\phi=0,\qquad \al^{-1}=\frac{2\pi
i}{\tau-\bar{\tau}},
\end{gather*} and ${\mathcal Im}(\tau)$ is the area of
parallelogram of periods. We give two representations of the Green
function $\phi$ and prove their equivalence using the same technique
as for the Kronecker series described in~\cite{We}.

A naive elliptic solution of (\ref{q111}) on $\ti W$ is obtained by
averaging (\ref{q10}) over the lattice
$\Gamma=\mZ\oplus\tau\mZ\subset\mC$:\footnote{We omit here and in
what follows the $\bz$ dependence.} \begin{gather}\label{q12}
\phi(z,y)=c\sum\limits_{\ga\in\Gamma}\frac{1}{\sqrt{(\pi\al
y)^2+|z+\ga|^2}}. \end{gather} However the series diverges.  That is why we
consider its generalization \begin{gather}\label{q120} \clR(s,x,z,y)=
c\sum\limits_{\ga\in\Gamma}\frac{\chi(\ga,x)}{({(\pi\al
y)^2+|z+\ga|^2})^s}, \qquad \clR\left(\tfrac{1}{2},0,z,y\right)=\phi(z,y), \end{gather}
where \begin{gather*}
\chi(\ga,x)=e^{\al^{-1}(\ga\bar{x}-\bar{\ga}x)}
\end{gather*} is a character ${\mathbb Z}\times{\mathbb Z}\rightarrow {\mathbb
C}^*$ of the additive group $\Gamma$ and $s$, $x$ are complex
parameters. The characters are double-periodic
\begin{gather*}
\chi(\ga,x+1)=\chi(\ga,x),\qquad
\chi(\ga,x+\tau)=\chi(\ga,x),\qquad \ga\in\Gamma, \end{gather*} while the
series $\clR(s,x,z,y)$ are quasi-periodic
\begin{gather}
  \clR(s,x,z+1,y)=e^{\al^{-1}(x-\bar x)}\clR(s,x,z,y), \nonumber\\
  \clR(s,x,z+\tau,y)=e^{\al^{-1}(x\bar\tau-\bar x\tau)}\clR(s,x,z,y).\label{qper1}
\end{gather}
 The variable $x$ describes behavior of $\clR(s,x,z,y)$ on the lattice $\Gamma$.
 In other words, $x$ parameterizes the moduli space of line bundles on $\Sigma_\tau$.
Note that for $\clR e\,s>1$  the series in the r.h.s.\ of
(\ref{q120}) converges.
 The function
\begin{gather}\label{f1} \clR\left(\tfrac{1}{2},x,z,y\right)=
c\sum\limits_{\ga\in\Gamma}\frac{\chi(\ga,x)}{({(\pi\al
y)^2+|z+\ga|^2})^{\frac{1}{2}}}=\phi(x,z,y)
 \end{gather}
 is the formal solution of (\ref{q111}) with the quasi-periodicity conditions (\ref{qper1}).

Another  representation of the Green function can be obtained by the
Fourier transform. Def\/ine the delta-functions \begin{gather*}
\delta(y)=\int\limits_{-\infty}^{+\infty}dp e^{2\pi ipy}, \qquad
\delta^{(2)}(z,{\bar z})=\sum\limits_{\ga\in\Gamma}\chi(\ga,z).
\end{gather*} Then \begin{gather*}
\sum\limits_{\ga\in\Gamma}\chi(\ga+x,z)=\chi(x,z)\sum\limits_{\ga\in\Gamma}\chi(\ga,z)=
\chi(x,z)\delta^2(z,{\bar{z}})=\delta^2(z,{\bar{z}}), \end{gather*} Let
$\ti\phi$ be the Green function with the quasi-periodicity
(\ref{qper1}) \begin{gather*}
\p^2_y\ti\phi+(2\pi\al)^2\p_{z}\p_{\bz}\ti\phi=c\delta(y)\delta^{(2)}(z,{\bar
z}). \end{gather*} Expanding it in the Fourier harmonics we f\/ind
\begin{gather*}
\ti\phi=-\frac{c}{4\pi^2}\sum\limits_{\ga\in\Gamma}\int\limits_{-\infty}^{+\infty}dp
\frac{ e^{2\pi ipy}}{p^2+|\ga+z|^2}\chi(\ga+x,z).
\end{gather*} Integrating
over $p$ provides factor $\pi$ and leads to the following
expression: \begin{gather}\label{q1211}
\ti\phi(x,z,y)=-\frac{c}{4\pi}\sum\limits_{\ga\in\Gamma}\frac{1}{|\ga+x|}e^{-2\pi|\ga+x||y|}\chi(\ga+x,z).
\end{gather} It is worthwhile to note that the solution (\ref{q1211}) is well
def\/ined. Our goal is to f\/ind interrelations between (\ref{q1211})
and~(\ref{f1}).

Consider a generalization of (\ref{q1211}) \begin{gather}\label{t}
I(s,x,z,y)=2c\pi^sy^{s-\oh}\sum\limits_{\ga\in\Gamma}
\frac{K_{s-\oh}(2\pi |y||\ga+x|)}{|\ga+x|^{s-\oh}} \chi(\ga+x,z).
\end{gather} Here $K_\nu$ is the Bessel--Macdonald function
\begin{gather*}
 K_{\nu}(2\pi yz)=\frac{\G(\nu+\oh)(z)^\nu}{2(\pi y)^\nu\G(\oh)}
 \int\limits_{-\infty}^{+\infty}dp\frac{ e^{2\pi ipy}}{(p^2+z^2)^{\nu+\oh}}.
 \end{gather*}
The function $I(s,x,z,y)$ is the Green function for the
pseudo-dif\/ferential operator \begin{gather*}
\left(\p_y^2+4\al^2\pi^2\p_{z}\p_{\bz}\right)^s \end{gather*} on
$\mR\times\Si_\tau$ with the boundary conditions (\ref{qper1}).
Since
\begin{gather*}
K_{\oh}(x)=\sqrt{\frac{\pi}{2x}}e^{-x},
\end{gather*} we
conclude that for $s=1$ $I$ coincides with $\ti\phi$ (\ref{q1211})
up to constant.

We are going to establish a relation between (\ref{q120}) and
(\ref{t}), and in this way between (\ref{q12}) and (\ref{q1211}).
Let us prove that \begin{gather}\label{q121} I(s,x,z,y)=
c\sum\limits_{\gamma\in\Gamma} \int\limits_{-\infty}^\infty
dp\int\limits_0^\infty \frac{d{t}}{t}t^s e^{-t(p^2+|\ga+x|^2)+2\pi
ipy}\chi(\ga+x,z). \end{gather} In fact, using the integral representation
for the Gamma-function \begin{gather}\label{gam}
\Gamma(s)=\int\limits_{0}^{\infty}\frac{dt}{t}t^{s}e^{-t} \end{gather} and
taking the integral over $t$ in (\ref{q121}) we come to (\ref{t}).

The representation (\ref{q121}) is universal and can serve to def\/ine
$\clR$ (\ref{q120})

\begin{lemma}\label{lemma3.1}
The function $\clR(s,x,z,y)$ has a representation as the Fourier
integral \begin{gather}\label{fi} \clR(s,x,z,y)=\f1{\G(s)\chi(\ga,z)}
\int_{\infty}^{\infty}dk \, I\left(s,z,x,\frac{k}{\pi\al}\right)e^{-2\pi iky}.
\end{gather}
\end{lemma}

\begin{proof} Substitute in (\ref{fi}) $I(s,x,z,y)$ (\ref{q121}) and
take
 f\/irst integral over $k$. We come to the condition $p=\pi\al y$.
 Then using the integral representation for the Gamma-function (\ref{gam}) we obtain (\ref{q120}).
\end{proof}

\begin{remark}\label{remark3.1}
The series (\ref{t}) is a three-dimensional generalization of the
Kronecker series (see~\cite{We})
\begin{gather*}
K(x,x_0,s)=\sum\limits_\ga\chi(\ga,x_0)|x+\ga|^{-2s}.
\end{gather*} Using the
Poisson summation formula Kronecker proved that
\begin{gather*}
\Gamma(s)K(x,x_0,s)=\al^{1-2s}\Gamma(1-s) K(x_0,x,1-s)\chi(x,x_0)\,.
\end{gather*}
\end{remark}

Our purpose is to generalize this functional equation
 for the 3-dimensional case $\Sigma_\tau\times \mathbb{R}$.
It takes the following form.

\begin{lemma}\label{lemma3.2}
The function $I(s,x,z,y)$ satisfies the  functional equation:
\begin{gather}\label{q122}
I(s,x,z,y)=\chi(x,z)\pi^{-\frac{1}{2}}\al^{-2s+1}\int\limits_{-\infty}^{+\infty}
dk\, I\left(\frac{3}{2}-s,z,x,\frac{k}{\pi\al}\right)e^{-2\pi i ky}.
 \end{gather}
\end{lemma}

\begin{proof} Following \cite{We} we subdivide integral (\ref{q121})
into two parts
\begin{gather*}
I(s,x,z,y)=c\sum\limits_{\gamma\in\Gamma}\int\limits_{-\infty}^\infty
dp\int\limits_0^T \frac{d{t}}{t}t^s e^{-t(p^2+|\ga+x|^2)+2\pi
ipy}\chi(\ga+x,z)\\
\phantom{I(s,x,z,y)=}{}+
c\sum\limits_{\gamma\in\Gamma}\int\limits_{-\infty}^\infty
dp\int\limits_T^\infty \frac{d{t}}{t}t^s e^{-t(p^2+|\ga+x|^2)+2\pi
ipy}\chi(\ga+x,z),\qquad T\in \mathbb{R},\quad T>0.\nonumber
\end{gather*}

The second term is a well def\/ined function for all $s$. Consider the
f\/irst one. It is well known that for the series
\begin{gather*}
\Theta(t,x,x_0)=\sum\limits_{\gamma}e^{-t|x+\gamma|^2}\chi(\ga,x_0)
\end{gather*} the following functional equation holds:
\begin{gather*}
\Theta(t,x,x_0)=(\al t)^{-1}\Theta\big(\al^{-2}t^{-1},x_0,x\big)\chi(x_0,x).
\end{gather*}

The latter follows from the Poisson summation formula
 which states that the averaging of function over some lattice equals
 the averaging of its Fourier transform over the dual lattice.
 In the above case the functional equation appears after
 the Fourier transform for the Gauss integral. Then
\begin{gather*} 
\sum\limits_{\gamma\in\Gamma}\int\limits_{-\infty}^\infty
dp\int\limits_0^T \frac{d{t}}{t}t^s e^{-t(p^2+|\ga+x|^2)+2\pi
ipy}\chi(\ga+x,z) \nonumber\\
\qquad{} = \sum\limits_{\gamma\in\Gamma}\int\limits_{-\infty}^\infty
dp\int\limits_0^T \frac{d{t}}{t}t^s
e^{-tp^2-\al^{-2}t^{-1}|\ga+z|^2+2\pi ipy}\chi(\ga+z,x)\chi(x,z)(\al
t)^{-1}\nonumber\\
\qquad{} \overset{\substack{\text{integrating}\\ \text{over $p$}}}{=}
\sum\limits_{\gamma\in\Gamma}\int\limits_0^T \frac{d{t}}{t}t^s
e^{-\pi^2{y^2}{t}^{-1}-\al^{-2}t^{-1}|\ga+z|^2}\chi(\ga+z,x)\chi(x,z)(\al
t)^{-1}\sqrt{\frac{\pi}{t}}\nonumber\\
\qquad{} \overset{\substack{\text{making} \\ \text{substitution} \\ \text{$\al^{-2}t^{-1}\rightarrow t$}}}{=}
\sum\limits_{\gamma\in\Gamma}\int\limits_{\al^{-2}T^{-1}}^\infty
\frac{d{t}}{t}t^{\frac{3}{2}-s} \sqrt{\pi}\al^{2-2s}e^{-t((\pi\al
y)^2+|\ga+z|^2)}\chi(\ga+z,x)\chi(x,z).
\end{gather*}
 Let $T=\al^{-1}$. Then
\begin{gather}  I(s,x,z,y)=
c\sum\limits_{\gamma\in\Gamma}\int\limits_{\al^{-1}}^\infty
\frac{d{t}}{t}t^{\frac{3}{2}-s} \sqrt{\pi}\al^{2-2s}e^{-t((\pi\al
y)^2+|\ga+z|^2)}\chi(\ga+z,x)\chi(x,z)\nonumber\\
\phantom{I(s,x,z,y)=}{}+
c\sum\limits_{\gamma\in\Gamma} \int\limits_{-\infty}^\infty
dp\int\limits_{\al^{-1}}^\infty \frac{d{t}}{t}t^s
e^{-t(p^2+|\ga+x|^2)+2\pi ipy}\chi(\ga+x,z).\label{q1213}
\end{gather}
 The proof follows
from (\ref{q1213}). One should only substitute $I(s,x,z,y)$ from
(\ref{q1213}), into (\ref{q122}). Formula (\ref{q1213}) represents
$I$ as the sum of two terms. Direct evaluation shows that the f\/irst
(of two) term from the l.h.s.\ of (\ref{q122}) equals to the second
one from the r.h.s.\ and vice versa.
\end{proof}

From Lemmas~\ref{lemma3.1} and~\ref{lemma3.2} we come to the main result of this
section
\begin{gather*}
\fbox{$
\clR(\frac{3}{2}-s,x,z,y)=\frac{\sqrt\pi\al^{2s-1}}{\G(\frac{3}{2}-s)}I(s,x,z,y).
$} \end{gather*} Now put  $s=1$.
Then one can see that well-def\/ined series \begin{gather}\label{phi}
\pi\sum\limits_{\gamma\in\Gamma}\chi(\ga+x,z)\frac{e^{-2\pi
|y||\ga+x|}}{|\ga+x|} \end{gather} describes the analytic continuation of the
divergent series
\begin{gather*}
\pi\sum\limits_{\gamma\in\Gamma}\chi(\ga,x)\frac{1}{\sqrt{(\pi\al
y)^2+|\ga+z|^2}}.
\end{gather*}

We use (\ref{q1211}) as the Green function.
Then
\begin{gather}\label{cA}
A_z(z,\bar{z},y,x)=-\frac{ic}{4\pi}\frac{1}{\pi^2\al^2}\hbox{sgn}(y)
\sum\limits_{\ga\in\Gamma}\frac{1}{\ga+x}e^{-2\pi|\ga+x||
y|}\chi(\ga+x,z),
\\
\hbox{sgn}(y)=1\quad \hbox{for}\quad y\geq 0,\qquad
\hbox{sgn}(y)=-1\quad \hbox{for}\quad y<0.\nonumber
\end{gather} Notice that the jump of $A$
(while coming through $y=0,\ z=0$) is obviously def\/ined by the jump
of $\hbox{sgn}(y)$.

\begin{remark}
Note that~(\ref{cA}) is a formal solution of the Bogomolny equation.
For $x\neq 0$ it is not a connection of a line bundle over
$\Si_\tau$ due to its monodromies similar to~(\ref{qper1}). We will
use this solution in next section to def\/ine a genuine connection for
higher ranks bundles.
\end{remark}

In order to compare elliptic conf\/iguration with the rational we take
$x=0$. Then on the line $y=0$ the connection is proportional to
\begin{gather*}
A_z\sim\sum\limits_{\ga\neq
0}\frac{1}{\ga}\chi(\ga,z)=E_1(z)-\al^{-1}(z-\bar{z}),
\end{gather*} where
$E_1(z)=\p\ln\vth(z)$ is the so-called f\/irst Eisenstein series and
 $\vth(z)$ is the theta-func\-tion~(\ref{theta}). $E_1(z)$ has a simple pole at $z=0$ with
$\hbox{Res}_{z=0}E_1(z)=1$ and the connection $A_z$ is
double-periodic. In terms of~(\ref{q3}) the gauge transformation $h$
compensating the jump of the connection is given by an integer power
of theta function $\vartheta^m(z)$, $m\in {\mathbb Z}$. Thus
\begin{gather*}
\p\log h=\p\log \vartheta^m(z)=mE_1(z). \end{gather*}

\section{Arbitrary rank case}\label{section4}

Here we describe modif\/ication of vector bundles of an arbitrary
rank. First, we repeat arguments of~\cite{KW} and justify the
choice $M$ in (\ref{mm}). As before, we consider
$\PSLN=G_{ad}$-bundles.

Near the singular point $\vec x^0$ the bundle $V$ is splited in a
sum of line bundles. Using the solu\-tion~(\ref{q10}) for a line
bundle we take the Higgs f\/ield near the singularity in the form
\begin{gather*}
\phi=\frac{i}{2\sqrt{y^2+z\bz}}\di(m_1,\ldots,m_N).
\end{gather*}
It follows from (\ref{q101}) that $A_z$ undergoes a discontinuous
jump at $y=0$ \begin{gather}\label{awga}
A^+_{z}-A^-_{z}=\frac{i}z\,\di(m_1,m_2,\ldots,m_N). \end{gather} To get rid
of the singularity of $A$ at $z=0$, as in the Abelian case, one can
perform the singular gauge transform $\Xi$ that behaves near $z=0$
as (\ref{gtr})
\begin{gather*}
\Xi=\di\big(z^{-m_1},z^{-m_2},\ldots,z^{-m_N}\big). \end{gather*}
 Assume that $\vec
m$ belongs to the weight lattice $\vec m\in P$. It means that $\Xi$
is inverse to the  cocharac\-ter~$\ga_{ad}$ of $\PSLN$ ($\ga_{ad}\in
t(G_{ad})=P^\vee\sim P$~(\ref{tb})). As it was explained before, the
modif\/ied bundle $V_+$ can not be lifted to a $\SLN$ bundle. On the
other hand, if $\vec m=(m_1,\ldots,m_N)$ belongs to the root lattice
$Q$ (\ref{tb}), then $\Xi^{-1}=\bar \ga$ and there is no obstruction
to lift $V_+$ to an $\SLN$ bundle. Note that (\ref{awga}) describes
the af\/f\/ine group $\bar W_a$ (\ref{q25}) action in the former case
and the af\/f\/ine group $W_a$ (\ref{sh}) action in the latter case.
From f\/ield-theoretical point of view it is an action of the t'Hooft
operator on $A_z$ (see~(\ref{a})).

 If $N=pl,$ $(l\neq 1,N)$
one can consider the intermediate situation and $\ga_{G_l}$
 (\ref{gl1}).
 It means that $\vec m\in t(^LG_l)\sim\G(G_p)$. This embedding provides the
modif\/ication that allows the $\PGLN$-bundle to lift to the
$G_l=\SLN/\mZ_l$-bundle  but not to a $\SLN$ bundle.
  In this way the monopole charges are related to  the characteristic classes of bundles.

One can use the maps to the Cartan subgroups  of the solution
$\phi(z)$ for a line bundle over $\Si_\tau$ (\ref{phi}) with $x=0$
\begin{gather*}
\phi\to \phi\cdot\di(m_1,m_2,\ldots,m_N). \end{gather*}
Unfortunately, in this case $V$ being restricted on $\Si_\tau$
 is splitting globally over $\Si_\tau$ and def\/ines an unstable
bundle, though it allows one  to describe its modif\/ications.

There exists a map of  $\phi(z,y,x)$ and $A_z(z,y,x)$ with $x\neq 0$
to a non-semisimple elements of $\sln$
\begin{gather*} 
\left(
              \begin{array}{cccc}
                0 & k_1\phi(z,y,x_1) & \dots & k_{N-1}\phi(z,y,x_{N-1}) \\
                0& 0 & \ldots & 0 \\
                \vdots & \ldots & \ddots & \vdots \\
                0 & \dots & \ldots & 0 \\
              \end{array}
            \right),
\\
\left(
              \begin{array}{cccc}
                0 & k_1A_z(z,y,x_1) & \dots & k_{N-1}A_z(z,y,x_{N-1}) \\
                0& 0 & \ldots & 0 \\
                \vdots & \ldots & \ddots & \vdots \\
                0 & \dots & \ldots & 0 \\
              \end{array}
            \right).
\end{gather*}
Since these matrices commute they are solutions of the matrix
equation (\ref{q6}). The connection has a jump at $y=0$. The bundle
is characterized by the diagonal monodromy  matrices (\ref{14})
\begin{gather*}
\rho_a=\di(a_1,a_2,\ldots,a_N),\qquad
\rho_b=\di(b_1,b_2,\ldots,b_N),
\end{gather*}
 where
 \begin{gather*} 
a_1=\prod_{j=1}^{N-1}\si_j^{\f1{N}},\qquad a_2=a_1\si_1^{-1}, \qquad a_N=a_1\si_{N-1}^{-1},
\\
b_1=\prod_{j=1}^{N-1}\varsigma_j^{\f1{N}},\qquad b_2=a_1\varsigma_1^{-1},\qquad
b_N=a_1\varsigma_{N-1}^{-1},
\end{gather*}
 $\sigma_j=\exp(\al^{-1}(x_j-\bar
x_j))$, $\varsigma_j=\exp(\al^{-1}(x_j\bar\tau-\bar x_j\tau
))$.
 Note that they are $y$-independent. Moreover,
the singular gauge transform, leading to a continues solution of the
Bogomolny equation, belongs to the upper nilpotent subgroup and in
this way does not change the topological type of the bundle.

Now we describe non-diagonal modif\/ications $\Xi$ of a
$\PGLN$-bundles over $\Si_\tau$. We do not know solutions of the
Bogomolny equation in this case, and only can assert that the
modif\/ication
``kill the jump'' of  $A_z$ at $y=0$:
 \begin{gather*}
 \Xi^{-1}\p_z\Xi=A_z^+-\Xi^{-1}A_z^-\Xi.
 \end{gather*}

We use the global description of a bundle $E$
 in terms of the transition matrices $\rho_a$, $\rho_b$ (\ref{14})
  using the approach of~\cite{LOZ1}.
 Let
 \begin{gather}\label{bc0}
 \rho_a={\rm Id}_N,\qquad \rho_b=\bfe\,{-u},\qquad (u=\di (u_1,u_2,\ldots,u_N))
 , \qquad {\bf e}(a)=\exp (2\pi i a).
 \end{gather}
The group commutator of these matrices is ${\rm Id}_N$. Thereby, $E$ can
be lifted to a $\SLN$-bundle.

 Def\/ine a modif\/ication $\Xi$ of $E$ to the bundle $\ti E$ with the transition
 matrices (\ref{qla}). Then $\Xi$ should intertwine the transition matrices
\begin{gather}\label{8.12}
\Xi(z+1,\tau)= \clQ\times \Xi(z,\tau), \\
\label{8.13}
\Xi(z+\tau,\tau)=\Lambda(z,\tau)\times \Xi(z,\tau) \times{\rm diag}
({\bf e}(u)).
\end{gather}
 The matrix $\Xi(z)$ degenerates at $z=0$ and we
assume that it has a simple pole. These conditions f\/ix $\Xi(z)$. It
can be expressed in terms of the theta-functions with
characteristics
\begin{gather*}
\Xi_{kj}(z, u_1,\ldots,u_N;\tau)=\frac{
\theta{\left[\begin{array}{c}
\frac{k}N-\frac12\\
\frac{N}2
\end{array}
\right]}(z-Nu_j, N\tau ) } { \theta^{\f1{N}}(z, \tau )}, \end{gather*} where
\begin{gather*}
\theta{\left[\begin{array}{c}
a\\
b
\end{array}
\right]}(z , \tau ) =\sum_{j\in \Bbb Z} \exp\,2\pi
i\left((j+a)^2\frac\tau2+(j+a)(z+b)\right). \end{gather*} The
quasi-periodicity properties (\ref{8.12}), (\ref{8.13}) follow from
the properties of the theta-functions
\begin{gather*}
\theta{\left[\begin{array}{c}
a\\
b
\end{array}
\right]}(z+1 , \tau )={\bf e}(a) \theta{\left[\begin{array}{c}
a\\
b
\end{array}
\right]}(z  , \tau ), \\
 \theta{\left[\begin{array}{c}
a\\
b
\end{array}
\right]}(z+a'\tau , \tau ) ={\bf e}\left(-{a'}^2\frac\tau2
-a'(z+b)\right) \theta{\left[\begin{array}{c}
a+a'\\
b
\end{array}
\right]}(z , \tau ).
\end{gather*}

This modif\/ication has the type
$(\frac{N-1}{N},-\f1{N},\ldots,-\f1{N})$. The modif\/ication that
allows to lift $\ti E$ to $\GLN$- bundle is
\begin{gather*}
\Xi_1(z)=h(z)\Xi(z)= \theta{\left[\begin{array}{c}
\frac{k}N-\frac12\\
\frac{N}2
\end{array}
\right]}(z-Nu_j, N\tau ), \end{gather*}
 where the gauge transformation $h$
is the diagonal matrix \begin{gather*}
 h(z)=\theta^{\f1{N}}(z, \tau
){\rm Id}_N.
\end{gather*} This modif\/ication intertwine the boundary conditions (\ref{bc0})
with \begin{gather*}
 \rho_a=\clQ,\qquad \rho_b=\ti\La,\qquad \ti \La=e^{-2\pi
i(\frac{z}N+\frac{\tau}{2N})}\La. \end{gather*} The last transformation
belongs to $\GLN$. Moreover, it can be proved that
\begin{gather*}
\det\left[\frac{\Xi_1(z, u_1,\ldots,u_N;\tau)} {i\eta(\tau)}\right]=
\frac{\vth(z)}{i\eta(\tau)}\prod\limits_{1\leq k<l\leq N}
\frac{\vth(u_l-u_k)}{i\eta(\tau)},
\end{gather*} where
$\eta(\tau)=q^{\frac{1}{24}}\prod_{n>0}(1-q^n)$ is the Dedekind
function $(q=\exp 2\pi i\tau)$ and $\vth(z)$ is the theta-function~(\ref{theta}).  Since $\vth(z)$  has a simple pole in $\Si_\tau$ the
bundle $\ti E$ is a $\GLN$-bundle of degree one. This modif\/ication
provides the Symplectic Hecke correspondence between the elliptic
Calogero--Moser system and the Elliptic Top.

Now consider the modif\/ication of the trivial bundle $E$ with the
transition matrices (\ref{bc0}) to the $\ti E=E_l$ (\ref{ic}),
(\ref{ic1}),
 where $\bfu_p=(\ti u_1,\ti u_2,\ldots,\ti u_p)$ is the moduli of the modif\/ied bundle.
The modif\/ication takes the form
\begin{gather*}
\Xi_{kj}(z, \tau) =
\frac{ \theta{\left[\begin{array}{c}
\frac{k}l-\frac12\\
\frac{l}2
\end{array}
\right]}(z-l\ti u_{i}), l\tau ) } { \theta^{\f1{l}}(z, \tau
)},\qquad (j=mp+i,\ m=0,\ldots,l-1).
\end{gather*} As it was explained in Section~\ref{section2} the modif\/ied bundle can be lifted to $G_l=\SLN/\mZ_l$-bundle, but
not to $\SLN$-bundle.

\appendix
\section[$\SLN$ and $\PSLN$ \cite{OV,Bo}]{$\boldsymbol{\SLN}$ and $\boldsymbol{\PSLN}$ \cite{OV,Bo}}

The group $\SLN$ is an universal covering of $\PSLN$ with the center
$\mZ_N=\mZ/N\mZ$ \begin{gather}\label{a1} {\rm Id} \to\mZ_N\to\SLN\to\PSLN\to {\rm Id}. \end{gather}
Therefore $\pi_1(\PSLN)=\mZ_N$. The both groups have the same Lie
algebra $\gG$.

 \textbf{Roots and weights.} The Cartan subalgebra $\gH\subset\gG$ is a hyperplane in $\mC^N$
\begin{gather*}
\gH=\left\{\bfx=(x_1,\ldots,x_N)\in\mC^N\,|\,\sum_{j=1}^Nx_j=0\right\}.
\end{gather*} The simple roots $\Pi=\{\al_k\}$
\begin{gather*}
\al_1=e_1-e_2,\quad \ldots,\quad \al_{N-1}=e_{N-1}-e_N \end{gather*} form a basis in the
dual space $\gH^*$.
 Here $\{e_j\}$ $j=1,\ldots,N$ is a canonical basis in $\mC^N$.
 They generate the set of roots of type $A_{N-1}$
 \begin{gather*}
 R=\{(e_j-e_k),~j\neq k\}.
\end{gather*}
The root lattice $Q\subset\gH^*$ takes the form \begin{gather}\label{rl}
Q=\left\{\sum m_je_j\,|\,m_j\in \mZ,~\sum m_j=0\right\}. \end{gather}

We identify $\gH^*$ and $\gH$ by means of the standard metric on
$\mC^N$. Then the coroot system
\begin{gather*}
R^\vee=\left\{\al^\vee(R)=
\frac{2(\al^\vee,\be)}{(\be,\be)}\in\mZ~{\rm for~any~}\be\in R\right\}
\end{gather*}
coincides with $R$, and the coroot lattice $Q^\vee$ coincides with
$Q$.

The fundamental weights $\varpi_k$, $(k=1,\ldots,N-1)$ are dual to
the basis of simple coroots $\Pi^\vee\sim\Pi$
$(\varpi_k(\al^\vee_k)=\de_{kj})$
\begin{gather}
\label{fw1}
\varpi_j=e_1+\dots+e_j-\frac{j}{N}\sum_{l=1}^Ne_l,\\
 \varpi_1=\left(\frac{N-1}N,-\f1{N},\ldots,-\f1{N}\right),\quad
 \varpi_2=\left(\frac{N-2}N,\frac{N-2}N,\ldots,-\frac{2}{N}\right),   \quad
 \ldots,\nonumber\\
    \varpi_{N-1}=\left(\frac{1}N,\frac{1}N,\ldots,\frac{1-N}{N}\right) . \nonumber
\end{gather} In the basis of simple roots the fundamental weights are
\begin{gather*}
\varpi_k= \f1{N}[(N-k)\al_1+2(N-k)\al_2+\dots+ (k-1)(N-k)\al_{k-1}\\
\phantom{\varpi_k=}{} +k(N-k)\al_k+k(N-k-1)\al_{k+1}+\dots+k\al_{N-1}].
\end{gather*}
The fundamental weights generate the weights lattice
 \begin{gather}
 \label{wl}
 P\subset\gH^*, \qquad P=\left\{\sum_ln_l\varpi_l\,|\,n_l\in\mZ\right\},
\\
 P=\sum_{j=1}^Nm_je_j, \qquad m_j\in \f1{N}\mZ, \qquad m_j-m_k\in\mZ.\nonumber
\end{gather}
 The weight lattice is generated by $Q$ and the vector
\begin{gather*}
\varpi_1=e_1-\f1{N}\sum_{j=1}^Ne_j. \end{gather*}
 The weight lattice $P$ def\/ines representations of $\SLN$, while
$Q$ def\/ine representations of $\PSLN$.

The factor-group  $P^\vee/Q^\vee\,$ $(P^\vee\sim P)$ is the center
$\mZ_N$ of $\SLN$. On the other hand it can be identif\/ied with the
cyclic group  symmetry $e_j\to e_{j+1}$ mod$(N)$ of the extended
Dynkin graph $\Pi\cup (\al_0=e_N-e_1)$.

\textbf{Characters and cocharacters.}
Let $\bar\clT$ $(\clT_{ad})$ be a  Cartan torus in $\SLN$ ($\PSLN$).
Def\/ine the groups of characters\footnote{The holomorphic maps of
the tori to $\mC^*$ such that $\chi(xy)=\chi(x)\chi(y)$ for
$x,y\in\clT$.}
\begin{gather*}
\bar\G=\{\bar\chi(x)\}=\{\bar\clT\to\mC^*\}, \qquad
\G_{ad}=\{\chi_{ad}(x)\}=\{\clT_{ad}\to\mC^*\}.
\end{gather*} They can be
identif\/ied with lattice groups in $\gH^*$ as follows.
Let $\varpi_k$ be a basic weight and
$\phi=(\phi_1,\phi_2,\ldots,\phi_N)$, $\phi_k=\f1{2\pi i}\ln
x_k$. The functions
\begin{gather*}
\exp 2\pi i
(\varpi_k\phi),\qquad k=1,\ldots,N-1
\end{gather*}
generate a basis in $\bar\G$.
Similarly, for $\al_k\in \Pi$
\begin{gather*}
\exp 2\pi i
(\al_k,\phi)
,\qquad k=1,\ldots,N-1
\end{gather*} is a basis in $\G_{ad}$.
Thereby, we have
\begin{gather*}
\bar\G=P,\qquad \G_{ad}=Q.
\end{gather*}

Def\/ine the dual groups   of cocharacters  $t(\bar G)=\bar\G^*$ and
$t(G_{ad})=\G_{ad}^*$ as the maps
\begin{gather*} 
t(\bar
G)=\{\bar\ga=\mC^*\to\bar\clT\}, \qquad
t(G_{ad})=\{\ga_{ad}=\mC^*\to\clT_{ad}\}.
\end{gather*} In another way
\begin{gather*}
 t(\bar G)=\{\phi\in \gH\,|\,\bar\chi(e^{2\pi
i\phi})=1\},\qquad  t(G_{ad})=\{\phi\in \gH\,|\,\bar\chi_{ad} (e^{2\pi
i\phi})=1\}. \end{gather*}
These groups are the groups of the coweight and
coroot lattices
 \begin{gather}\label{tb} t(\bar G)=Q^\vee\sim
Q,\qquad t(G_{ad})=P^\vee\sim P.
\end{gather}
The center of $\bar\G=\SLN$
belongs to $\bar \clT$ and is identif\/ied with the factor-group
\begin{gather}
\label{center} \clZ(\SLN)=P^\vee/t(\bar G)\sim P^\vee/Q^\vee\sim \\
\phantom{\clZ(\SLN)=}{}\sim \pi_1(\PSLN)\sim t(G_{ad})/Q^\vee\sim P/Q=\mZ_N.\nonumber
\end{gather}

Let  $N=pl$, $(l\neq 1,N)$ and $\mZ_l\subset\mZ_N$ be a subgroup
of $Z_N$.
 Def\/ine the factor-group
\begin{gather*}
G_l=\SLN/\mZ_l.
\end{gather*}
 Then the center $\clZ(G_l)$  of $G_l$ is $\mZ_p$ and
$\pi_1(G_l)=\mZ_l$. Consider the groups of characters and
cocharacters of $G_l$
\begin{gather}
\label{gl} \G(G_l)=\{\chi_{G_l}\,:\,\clT(G_l)\to
\mC^*\}, \\
 \label{gl1}
t(G_l)=\{\ga_{G_l}\,:\,\mC^*\to\clT(G_l)\},
\end{gather}
$(\G^*(G_l)=t(G_l))$. They are lattices in $\gH^*$ and $\gH$
$Q\subset\G(G_l)\subset P$, $Q^\vee\subset t(G_l)\subset P^\vee$.
The lattice $\G(G_l)$ is generated by the root lattice $Q$ and the
vector $l\varpi_1$, while the lattice $t(G_l)$ is generated by the
root lattice $Q$ and the vector $p\varpi_1$ \begin{gather}\label{gg}
\G(G_l)=l\varpi_1\cup Q,\qquad t(G_l)=p\varpi_1\cup Q. \end{gather} The group
$\G(G_l)$ is the weight lattice of $G_l$ because  highest weights of
irreducible f\/inite-dimensio\-nal representations of $G_l$ belong to
$\G(G_l)$.

In terms of lattices the center  $\clZ(G_l)$ and $\pi_1(G_l)$ take
the form
\begin{gather*}
\clZ(G_l)\sim P^\vee/t(G_l)\sim
\G(G_l)/Q\sim\mZ_p, \\
\pi_1(G_l)\sim
t(G_l)/Q^\vee\sim P/\G(G_l)\sim\mZ_l.
\end{gather*} A subgroup $^LG_l\subset
\SLN$ is the Langlands dual to $G_l$ if
\begin{gather*}
t(^LG_l)\sim\G(G_l)\qquad (\G(^LG_l)\sim t(G_l)). \end{gather*}
 It
implies that
\begin{gather}
\label{cdg} \clZ(^LG_l)\sim \mZ_l, \\
\pi_1(^LG_l)\sim \mZ_p. \nonumber
\end{gather}
Therefore the dual group is
\begin{gather}
\label{dugr} ^LG_l=G_p.
\end{gather}
 In particular, $^L\SLN=\PGLN$.

\textbf{Af\/f\/ine Weil group.}
The af\/f\/ine Weyl group $W_a$ is a semidirect product $Q^\vee\rtimes
W$ of
 the Weyl group $W$ and the group $Q^\vee$. It acts on $\gH$ as
\begin{gather}\label{sh} x\to
x-\frac{2(\al,x)}{(\al,\al)}\al^\vee+k\al^\vee,\qquad k\in\mZ. \end{gather}
 Consider a semidirect product
 \begin{gather}\label{q25}
 \bar W_a=P^\vee\rtimes W.
 \end{gather}
  In particular,
 the shift operator
 \begin{gather}\label{shi}
 x\to x+\vec m,\qquad \vec m\in P^\vee
 \end{gather}
 is an element from $\bar W_a$. It follows from this construction that
 the factor group
 \begin{gather}\label{fwe}
 \bar W_a /W_a\sim P^\vee/Q^\vee\sim\clZ(\SLN).
 \end{gather}
Let again $N=pl$ and  def\/ine  a subgroup $W_a(G_l)$ of $\bar W_a$,
generated by shifts from $t(G_l)$ \begin{gather*}
 W_a(G_l)=t(G_l)\rtimes W.
 \end{gather*}
The factor group $W_a(G_l) /W_a$ is isomorphic to $t(G)/Q^\vee$ and
in this way
 to $\clZ(^LG_l)$.

\subsection*{Acknowledgements}

The  work   was  supported  by   grants
RFBR-09-02-00393,  RFBR-09-01-92437-$KE_a$ and
NSh-3036.2008.2. The work of A.Z.\ was also supported by the
``Dynasty'' fund, the President fund MK-5242.2008.2,
NWO-RFBR-05-01-80006 and RFBR-09-01-93106-$NCNIL_a$. The authors are
grateful for hospitality to the Max Planck Institute of Mathematics,
Bonn, where the most part of this work was done.

\pdfbookmark[1]{References}{ref}
\LastPageEnding

\end{document}